\documentclass[aps,nofootinbib,twocolumn]{revtex4}
\usepackage{graphicx,amsmath,hyperref}

\newcommand{\be}{\begin{equation}}
\newcommand{\ee}{\end{equation}}

\begin{document}
\title{ \Large Simple model for quantum general relativity from  loop quantum gravity}
     \author{Carlo Rovelli}
     \affiliation{Centre de Physique Th\'eorique de Luminy,
     \footnote{Unit\'e mixte de recherche du CNRS et des Universit\'es de Provence, de la M\'editerran\'ee et du Sud; affili\'e \`a la FRUMAN.}
      Case 907, F-13288 Marseille, EU
     }
\date{\small  \today}
\begin{abstract} \noindent 
New progress in loop gravity has lead to a simple model of  `general-covariant quantum field theory'.   I sum up the definition of the model in self-contained form, in terms accessible to those outside the subfield. I emphasize  its formulation as a generalized topological quantum field theory with an infinite number of degrees of freedom, and its relation to lattice theory. I list the indications supporting the conjecture that the model is related to general relativity and UV finite.
(This contribution has appeared also in the "Spanish Relativity Meeting (ERE2010)  Proceedings" in the open access Journal of Physics: Conference Series (JPCS))
\end{abstract}

\maketitle

\section{The Model} 

A simple model has recently emerged in the context of loop quantum gravity.  
It has the structure of a generalized topological quantum field theory (TQFT),  with an infinite number of degrees of freedom, local in sense of classical general relativity (GR). It can be viewed as an example of a ``general-covariant quantum field theory". It is defined as a function of two-complexes and may have mathematical interest in itself.  I present the model here in concise and self-contained form.

The model has emerged from the unexpected convergence of many lines of investigation, including canonical quantization of GR in Ashtekar variables \cite{Ashtekar86,Rovelli:1987df,Rovelli:1989za,Ashtekar:1992tm,Kaminski:2009fm}, Ooguri's \cite{Ooguri:1992eb} 4d generalization  of matrix models \cite{Brezin:1977sv,David:1985vn,Kazakov:1985ea,Ambjorn:1985az,Gross:1989vs}, covariant quantization of GR on a Regge-like lattice \cite{Engle:2007uq,Engle:2007qf,Engle:2007wy}, quantization of geometrical ``shapes" \cite{Barbieri:1997ks,Barrett:1997gw,Livine:2007vk,Freidel:2007py} and Penrose spin-geometry theorem \cite{Penrose2}.  The corresponding literature is intricate and long to penetrate.  Here I skip all `derivations' from GR, and, instead, list the elements of evidence supporting the conjectures that the transition amplitudes are finite and the classical limit is GR.

The model's dynamics is defined in Sec.\,\ref{model}. States and operators in Sec.\,\ref{hilbert} and\,\ref{penrose}.  Sec.\,\ref{support} reviews the evidence relating the model to GR, and some of its properties.

\section{Feynman rules}\label{model}

The model is defined assigning transition amplitudes $Z_{\cal C}(h_l)$ with $h_l\!\in\! SU2$, to two-complexes $\cal C$ with boundary.  

\begin{figure}[t]
\centerline{
\includegraphics[scale=0.15]{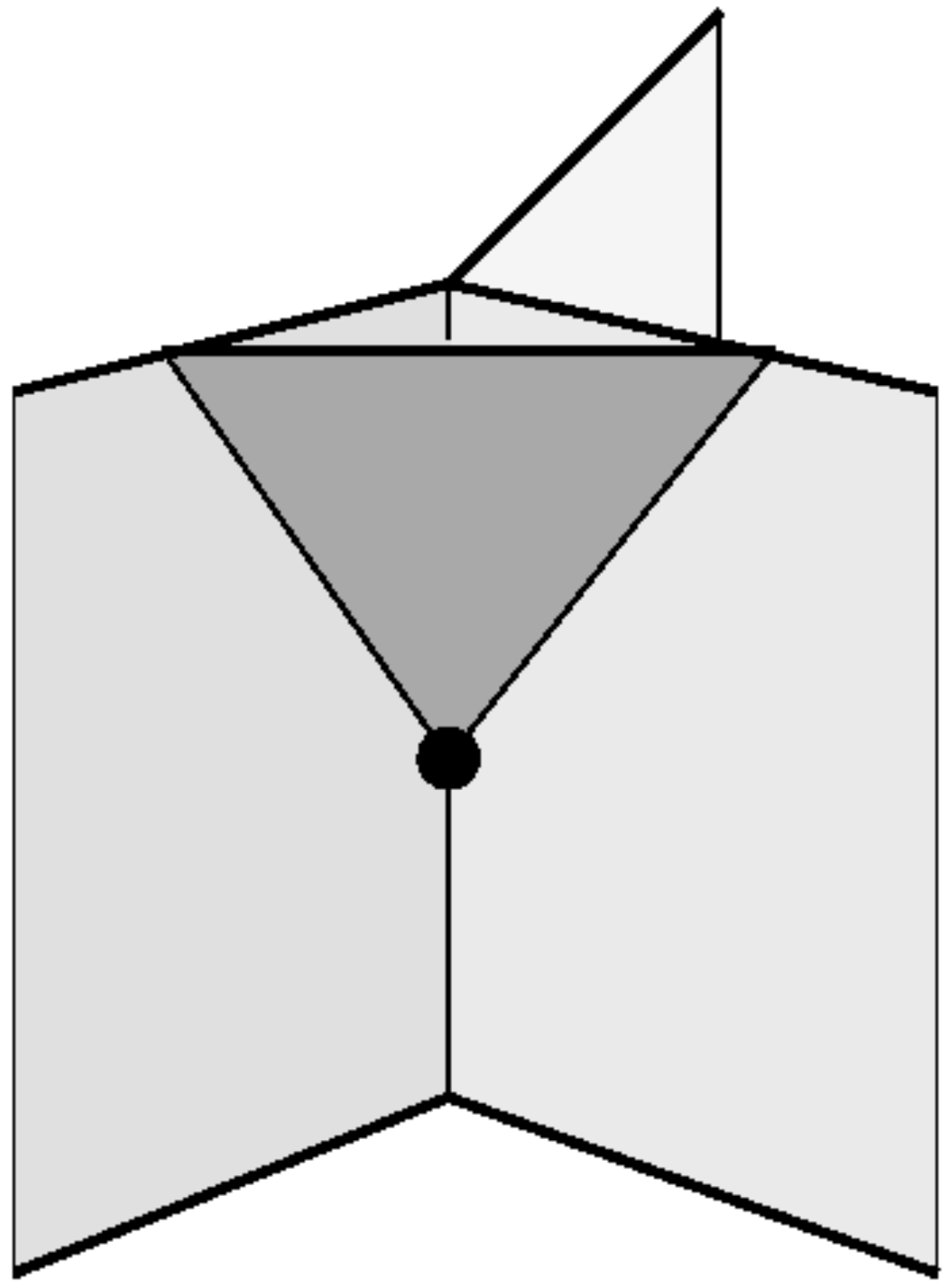}
\hspace{-5.6em}
\begin{picture}(40,30)
\put(20,23){\small $v$}
\put(25,34){\small $e$}
\put(2,22){\small $f$}
\put(12,57){\small $n$}
\put(25,67){\small $l$}
\end{picture}\hspace{5em}
}\caption{A two-complex with one bulk vertex.}\vspace{-1em}
\label{13}
\end{figure}
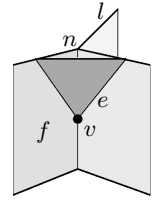

A two-complex (see Fig.1) is a finite set of $F$ elements $f$ (``faces"), $E$ elements $e$ (``edges"), and $V$ elements $v$ (``vertices"), equipped with a boundary relation $\partial$ associating an ordered couple of vertices $\{s_e,t_e\}$ (``source", ``target") to each edge and a cycle of edges $\{e_{1f},...,e_{nf}\}$ to each face. 
Its boundary is a (possibly disconnected) graph $\Gamma$, whose $L$ links $l$ are edges of $\cal C$ bounding a single face and whose $N$ nodes $n$ are vertices of $\cal C$ bounding (links and) a single internal edge.  $Z_{\cal C}(h_l)$ is defined as the integral obtained associating:
\begin{enumerate}\addtolength{\itemsep}{-2mm}
\item Two group integrations  to each internal edge (or one to each adjacent couple  \{internal edge, vertex\})\vspace{-5mm}
\be
\begin{picture}(35,25)
\thicklines
\put(-7,-11){\tiny $g'$}
\put(22,16){\tiny $g$}
\put(0,-6){\line(1,1){20}}
\put(10,-1){\tiny $e$}
\end{picture}\longmapsto\ \ 
\int_{SL2C}dg_{e s_e}\int_{SL2C}dg_{et_e}
\label{un}
\ee
\item A group integration to each couple of adjacent \{face, internal edge\} 
\vspace{-2mm}
\be
\begin{picture}(35,25)
\put(0,-6.2){\line(-1,0){10}}
\put(20.2,14){\line(-1,1){10}}
\thicklines
\put(0,-6){\line(1,1){20}}
\put(7,-3){\tiny $e$}
\put(-8,8){\tiny $f$}
\put(2,9){\tiny $h_{e\!f}$}
\end{picture}\longmapsto\ \ 
\int_{SU2}dh_{e\!f}\; \chi^{j_f}(h_{e\!f})
\label{su2int}
\ee
$\chi^{j}(h)$ is the spin-$j$ $SU2$ character of $h$.
\item A sum  to each face $f$
\vspace{-2mm}
\be
\hspace{3em}
\begin{picture}(25,25)
\thicklines
\put(-10.1,-6){\line(0,1){20}}
\put(-0.1,-6){\line(-1,0){10.3}}
\put(0,24){\line(1,0){10}}
\put(-10,14){\line(1,1){10}}
\put(20,14){\line(-1,1){10}}
\put(0,-5.9){\line(1,1){20}}
\put(11,0){\tiny $h_{e\!f}$}
\put(-1,8){\tiny $f$}
\put(0,-11){\tiny $g'$}
\put(22,11){\tiny $g$}
\end{picture}
\longmapsto\ \sum_{j_{\!f}}d_{j_{\!f}}\,\chi^{\scriptscriptstyle\gamma (j_{\!f}+1),j_{\!f}}\!\Big(\!\prod_{e\in\partial f}g_{e\!f}^{\epsilon_{l\!f}}\!\Big).
\label{sei}
\ee
where $
g_{ef}:= g_{es_e} h_{e\!f}g^{-1}_{et_e}
$ for internal edges, and $g_{e\!f}$ = $h_l\!\in\! SU2$ for boundary edges. $d_j$ is $(2j+1)$. $\chi^{p,k}(g)$ is the $SL2C$ character in the unitary representation with (continuous and discrete) Casimir eigenvalues $p$ and $k$. $\epsilon_{e\!f}\!=\!\pm1$ according to whether the orientations (defined by $\partial$) of the edge $e$ and the face $f$ are consistent or not.   $\gamma$ is a fixed real parameter called Barbero-Immirzi parameter. 
\item At each vertex, one of the integrals $\int_{SL2C} dg_{ev}$ in \eqref{un} (which is redundant) is dropped. 
\end{enumerate}
The resulting amplitude can be written compactly as \vspace{-2mm}
\begin{eqnarray}
&&Z_{\cal C}(h_l)= \int_{(SL2C)^{\scriptscriptstyle 2(E-L)-V}}dg_{ve}\int_{(SU2)^{ \scriptscriptstyle  {\cal V}-L}}dh_{e\!f}\;
\nonumber\\
\label{int1}
&&\hspace{1em}  \sum_{{j_{\!{}_f}}}  \prod_{f} d_{j_{\!{}_f}}\;
\chi^{\scriptscriptstyle\gamma ({j_{\!{}_f}}\!+1),{j_{\!{}_f}}}\!\Big(\!\prod_{e\in\partial f}g_{e\!f}^{\epsilon_{l\!f}}\!\Big) \prod_{e\in\partial f}\chi^{j_{\!{}_f}}\!(h_{e\!f})
\end{eqnarray}
where  $\cal V$ is the sum of the valences of all the faces. This completes the definition of the model. 

For a two-complex without boundary, \eqref{int1} reduces to the ``partition function"
\begin{eqnarray}
\label{int2}
&&Z_{\cal C}= \int_{(SL2C)^{2E-V}}dg_{ve}\int_{(SU2)^{\cal V}}dh_{e\!f}\;
 \sum_{{j_{\!{}_f}}}
 \prod_{f} d_{j_{\!{}_f}}
  \\ \nonumber
&&\hspace{1em}   
\chi^{\scriptscriptstyle\gamma ({j_{\!{}_f}}\!+1),{j_{\!{}_f}}}\!\Big(\!\prod_{e\in\partial f}(g_{es_e} h_{e\!f} g^{-1}_{et_e})^{\epsilon_{l\!f}}\!\Big) \prod_{e\in\partial f}\chi^{j_{\!{}_f}}\!(h_{e\!f}).
\end{eqnarray}
(The sum defining the $SL2C$ character converges because the $SU2$ integral reduces it to a finite subspace.)  A formulation more similar to the one common in the literature is in Sect.\,\ref{spinfoams} (the one above is related to \cite{Geloun:2010vj}). 

Section \ref{hilbert} clarifies in which sense the $Z_{\cal C}(h_l)$ define a general covariant QFT, and Section \ref{support} clarifies the relation with GR, and how these transition amplitudes can be used to compute physical quantities such as graviton's $n$-points functions or the evolution of a classical spacetime. Before going into this, however, I anticipate some comments on the intuitive physical interpretation of these quantities.

There are two related but distinct physical interpretations of the above equations, that can be considered. The first is as a concrete implementation of Misner-Hawking intuitive ``sum over geometries" 
\be
Z=\int_{\rm Metrics/Diff}\ Dg_{\mu\nu}\ e^{\frac{i}{\hbar}S[g_{\mu\nu}]}.
\label{sog}
\ee
As we shall see, indeed, the integration variables in  \eqref{int2} have a natural interpretation as 4d geometries (Sect. \ref{spinfoams}), and the integrand approximates the exponential of the Einstein-Hilbert action $S[g_{\mu\nu}]$ in the semiclassical limit (Sect.\ref{support}). Therefore \eqref{int2} gives a family of approximations of \eqref{sog} as the two-complex is refined.  But there is a second interpretation, compatible with the first but more interesting: the transition amplitudes \eqref{int1}, formally obtained  sandwiching the sum over geometries \eqref{sog} between appropriate boundary states, 
can be interpreted as terms in a generalized perturbative Feynman expansion for the dynamics of quanta of space (Sect.\;\ref{spinnetworks}). In particular, \eqref{int1} implicitly associates a vertex amplitude (given explicitly below in \eqref{va}) to each vertex $v$: this is the general-covariant analog for GR of the QED vertex amplitude 
\be
   \raisebox{-.5cm}{\includegraphics[scale=0.4]{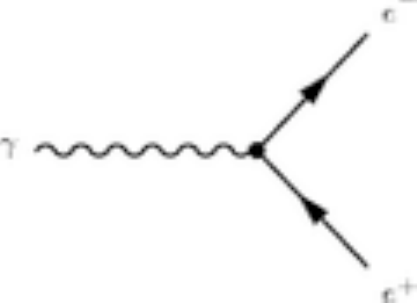}}
   =
   e\ \gamma_\mu^{AB}\ \delta (p_{1}\!\!+\!p_{2}\!\!+\! k ). 
   \label{qed}
\ee
Therefore the transition amplitudes \eqref{int1} are a general covariant and background independent analog of the Feynman graphs. These remarks about interpretation should become more clear in the last section. 

The model has a euclidean version \cite{Engle:2007qf,Freidel:2007py}, obtained replacing $SL2C$ with $SO4$, and can be written in a (euclidean or lorentzian) quantum deformed version, obtained by replacing  $SO4$ and $SL2C$ with their $q$ deformation (see \cite{Noui:2002ag}). The $q$-deformed version has not yet been sufficiently studied, but one might expect it to correspond to the the inclusion of a cosmological constant and its transition amplitudes \eqref{int1} to be finite for appropriate values of $q$. 

\section{TQFT on manifolds with defects}\label{hilbert}

Atiyah has provided a compelling definition of a general covariant QFT, by giving axioms for topological quantum field theory (TQFT) \cite{Atiyah:1988fk,Atiyah:1990uq}.  In Atiyah scheme, a 4d TQFT is defined by the cobordisms between 3d manifolds.   To each compact 3d manifold $M_3$ without boundaries is associated a finite dimensional Hilbert space ${\cal H}_{M_3}$, and to each 4d manifold $M_4$ with boundary  $\partial {M_4}$ is associated a state $\psi_{M_4}\in{\cal H}_{\partial {M_4}}$. These satisfy natural composition axioms. 

The model defined by \eqref{int1} belongs to a simple generalization of Atiyah's TQFT, where: (i) boundary Hilbert spaces are not necessarily finite dimensional; (ii) 4d manifolds are replaced by two-complexes; (iii) 3d manifolds are replaced by graphs \cite{Baez:1997zt,Baez:1999sr,Crane:2010fk}.  Graphs bound two-complexes in the same manner in which 3d manifolds bound 4d manifolds.

Consider a graph $\Gamma$, namely a set of $L$ elements $l$ called ``links" and  $N$ elements $n$ called ``nodes", and a boundary relation $\partial$  associating to each link an ordered couple of nodes $\partial l=\{s_l,t_l\}$. Associate to each graph $\Gamma$ the Hilbert space
\be
 {\cal H}_\Gamma= L_2[(SU2)^L/(SU2)^N]
\label{n-1}
\ee
where the $L_2$ is defined by the Haar measure and the ``gauge" action of $(SU2)^N$ on the states $\psi(h_l)\in L_2[(SU2)^L]$ is 
\be
\psi(h_l) \to \psi(V_{s_l}h_l V_{t_l}^{-1}),\hspace{2em}ÊV_n\in (SU2)^N.
\label{gauge}
\ee
If $\cal C$ is a two-complex bounded by the (possibly disconnected) graph $\Gamma$, then \eqref{int1} defines a state in ${\cal H}_\Gamma$ which satisfies TQFT composition axioms \cite{Bianchi:2010fj}. Thus, the model defined above defines a generalized TQFT in the sense of Atiyah.%
\footnote{This generalization consists essentially in replacing manifolds $M$ by ``manifold with defects" $\tilde {M}$. A graph $\Gamma$ is related to a 3d manifold with defects $\tilde M_3$ as follows. Take a cellular decomposition $\Delta$ of a 
(say, topologically trivial) 3d manifold $M_3$. Then $\tilde M_3$ is constructed removing the 1-skeleton $\Delta_1$ of $\Delta$ from $M_3$, that is $\tilde M_3=M_3\!-\!\Delta_1$, and $\Gamma$ is identified with $\Gamma\!=\!\Delta_1^*$, the 1-skeleton of the dual complex. Notice that $\Gamma$ captures fully the fundamental group of $\tilde M_3$. Similarly, a two-complex $\cal C$ can be related to a 4d manifold with defects $\tilde M_4$ by $\tilde M_4\!=\!M_4-\Delta_2$ and ${\cal C}\!=\!\Delta^*_2$, namely removing the 2-skeleton of the cellular complex, and identifying $\cal C$ with the 2-skeleton of the dual complex.  Now, recall that in Regge gravity curvature is concentrated on defects with codimension 2, and the holonomy of the Levi-Civita connections on the \emph{flat} manifold with defects $M_n\!-\!\Delta_{n\!-\!2}$ captures entirely the geometry.  Manifolds with codimension-2 defects (or graphs and two-complexes) are this natural carriers of \emph{curved} Regge geometries.  In \cite{Bianchi:2009tj}, the space  ${\cal H}_\Gamma$ is precisely constructed as the quantization of a space of flat $SU2$ connections on $\tilde M_3$, or equivalently a space of Regge metrics where curvature is on the defects.} 

In the next section I show (following \cite{Bianchi:2010kx}) that the states in this boundary space have a natural interpretation as 3-geometries, thanks to a beautiful theorem by Penrose. 

\section{Penrose metric operator}\label{penrose}

The boundary Hilbert space \eqref{n-1} has a natural interpretation as a space of quantum metrics, that was early recognized by Roger Penrose.  The natural ``momentum" operator on $L_2[SU2]$ is the derivative operator
\be
L^i\psi(h)\equiv \left.i\frac{d}{dt}\, \psi(he^{it \tau_i})\right|_{t=0},
\ee
where $i=1,2,3$ labels a hermitian basis $\vec \tau=\{\tau_i\}$ in the $su2$ algebra. The gauge invariant operator
\be
G_{ll'}= \vec L_l\cdot \vec L_{l'}
\ee
where $\vec L_l=\{L_l^i\}$ is the derivative with respect to $h_l$ and  $s_l=s_{l'}:=n$, is well defined on ${\cal H}_\Gamma$ and coincides with Penrose's metric operator \cite{Major:1999mc}. Penrose spin-geometry theorem then gives states in ${\cal H}_\Gamma$ a consistent interpretation as quantized 3-geometries. The metric operator $G_{ll'}$ determines the angle between the links $l$ and $l$ at the node $n$   \cite{Penrose2,Moussouris:1983uq,Major:1999mc} (see Fig.2). The theorem states that these angles obey the dependency relations expected of angles in three dimensional  space.  A volume element associated to the node $n$ can be defined in terms of Penrose metric operator, using standard relations between metric and volume element \cite{Rovelli:1994ge}. For instance, for a 4-valent node $n$, bounding the links $l_1,...,l_4$ the volume operator $V_{n}$ is given by 
\be 
V^2_{n}= |\vec L_{l_1}\cdot(\vec L_{l_2}\times \vec L_{l_3})|;
\ee
gauge invariance \eqref{gauge} at the node ensures that this definition does not depend on which triple of links is chosen.  Analogously, the diagonal terms $G_{ll}$ of the metric determines the area element $A_l$ normal to the link $l$ by
\be 
A^2_{l}= \vec L_{l}\cdot \vec L_{l}.
\ee
The Area and Volume operators  $A_l$ and $V_n$ form a complete set of commuting observables in ${\cal H}_\Gamma$, in the sense of Dirac. The spectrum of both operators can be computed \cite{Rovelli:1994ge}; it is discrete and it has a minimum step between zero and the lowest non-vanishing eigenvalue. In the case of the area, this gap is 
\be
a_0=\frac{\sqrt 3}2. 
\ee

The orthonormal basis that diagonalizes the complete commuting commuting set of operators $A_l,V_n$ is called the \emph{spin-network} basis. This basis can be obtained  via the Peter-Weyl theorem. It is labelled by a spin $j_l$ for each link $l$ and an $SU2$ intertwiner $i_n$ for each node $n$  \cite{Rovelli:1995ac,Baez95a,Baez95aa}, and defined by 
\be
\psi_{\Gamma, j_l,i_n}(h_l)=\big\langle \otimes_l\,d_{j_l}\, D^{j_l}(h_l)\; \big |\; \otimes_n i_n\; \big\rangle {}_\Gamma
\label{sn}
\ee
where $D^{j_l}(h_l)$ is the Wigner matrix in the spin-$j$ representation and $\langle \cdot |\cdot  \rangle_\Gamma$ indicates the pattern of index contraction between the indices of the matrix elements and those of the intertwiners given by the structure of the graph.\footnote{Both tensor products live in 
\be
H_\Gamma\subset L_2[(SU2)^L]=\bigoplus_{j_l} \bigotimes_l\, {\cal V}_{j_l}\otimes{\cal V}_{j_l}=\bigoplus_{j_l} 
\bigotimes_n\bigotimes_{e\in \partial n}{\cal V}_{j_l}.
\ee
where ${\cal V}_j$ is the $SU2$ spin-$j$ representation space, here identified with its dual.} A $G$-intertwiner, where $G$ is a Lie group, is an element of a (fixed) basis of the $G$-invariant subspace of the tensor product $\otimes_l {\cal H}_{j_l}$ of irreducible $G$-representations ---here those associated to the links $l$ bounded by $n$.  Since the Area is the $SU2$ Casimir, the spin $j_l$ is easily recognized as the Area quantum number and $i_n$ is the Volume quantum number. 

\begin{figure}[t]
\centerline{\includegraphics[scale=0.3]{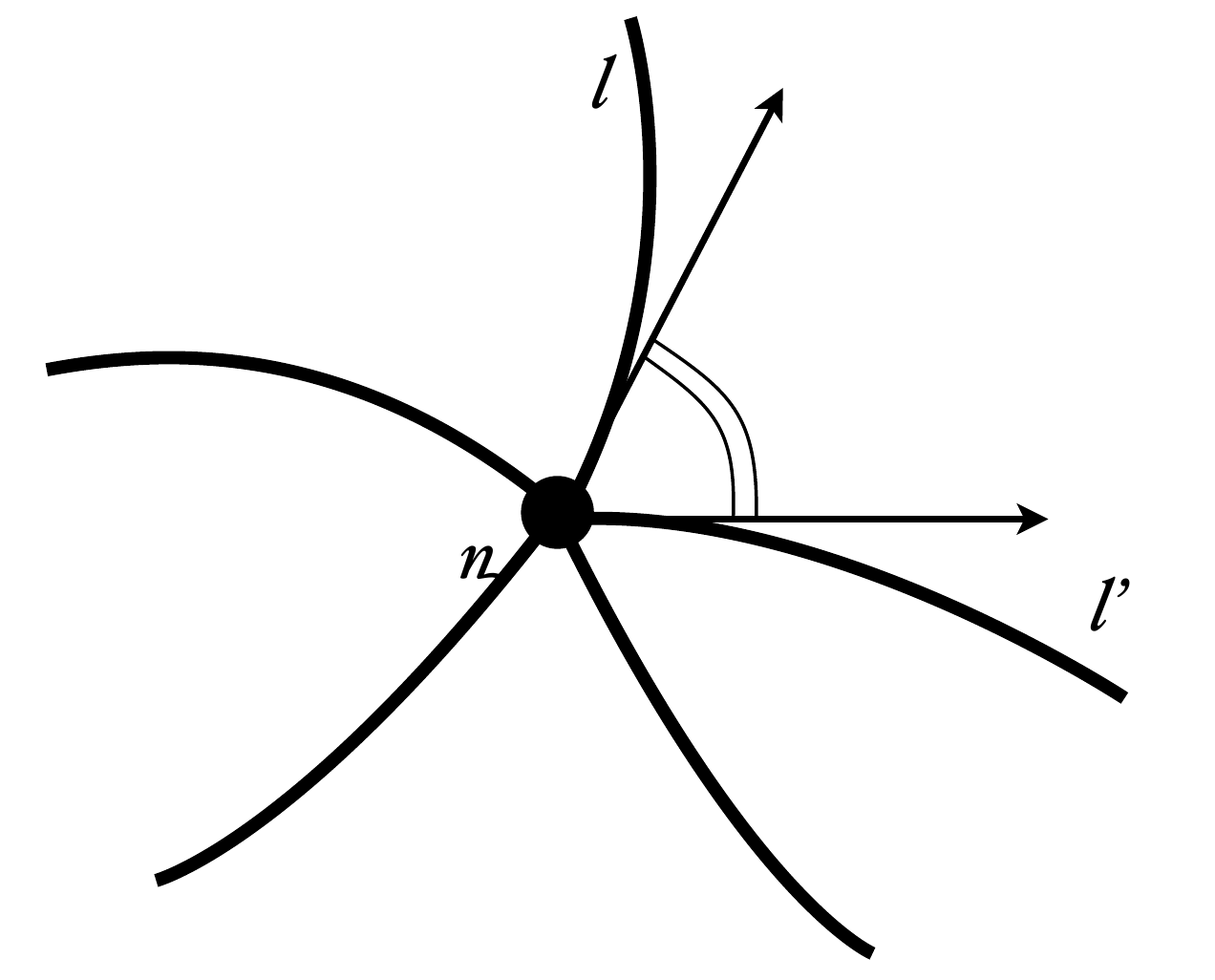}}
\caption{The angle defined by the Penrose metric operator on the graph.} 
\end{figure}

Coherent states in ${\cal H}_\Gamma$ have been studied by a number of authors and are particularly useful in applications \cite{Hall:2002, Ashtekar:1994nx,Thiemann:2002vj,Livine:2007vk,Bianchi:2009ky,Bianchi:2010gc,Freidel:2010aq,Rovelli:2010km,Freidel:2010bw} (see also \cite{Dittrich:2008ar,Oriti:2009wg,Bonzom:2009wm}).

\subsection{Spin networks as quantum 3-geometries}\label{spinnetworks}

The results above equip the boundary states of the model \eqref{int1} with a geometrical interpretation: the spin network state $\psi_{\Gamma, j_l,i_n}$ is interpreted as representing a granular space. Each node is a quantized ``chunk", or ``quantum" of space (see Fig.3); the graph gives the connectivity relations between these quanta; $i_n$ is the quantum number of the volume of the $n$'th quantum of space; and $j_l$ is the quantum number of the area of the elementary surface separating the adjacent nodes $s_l$ and $t_l$. 

Thus, the quantum states of the theory describe background-independent quantum excitations of the geometry of space.  Physical space is built up, or ``weaved up"  \cite{Ashtekar:1992tm} by such nets of atoms of space. 

As in classical GR \cite{Rovelli:1990ph,Rovelli:2001my}, and unlikely in ordinary field theory, in this theory localization is only relative to the field itself. In this sense, the theory is profoundly different from ordinary \emph{local} quantum field theory.

\begin{figure}[t]
\centerline{\includegraphics[scale=0.2]{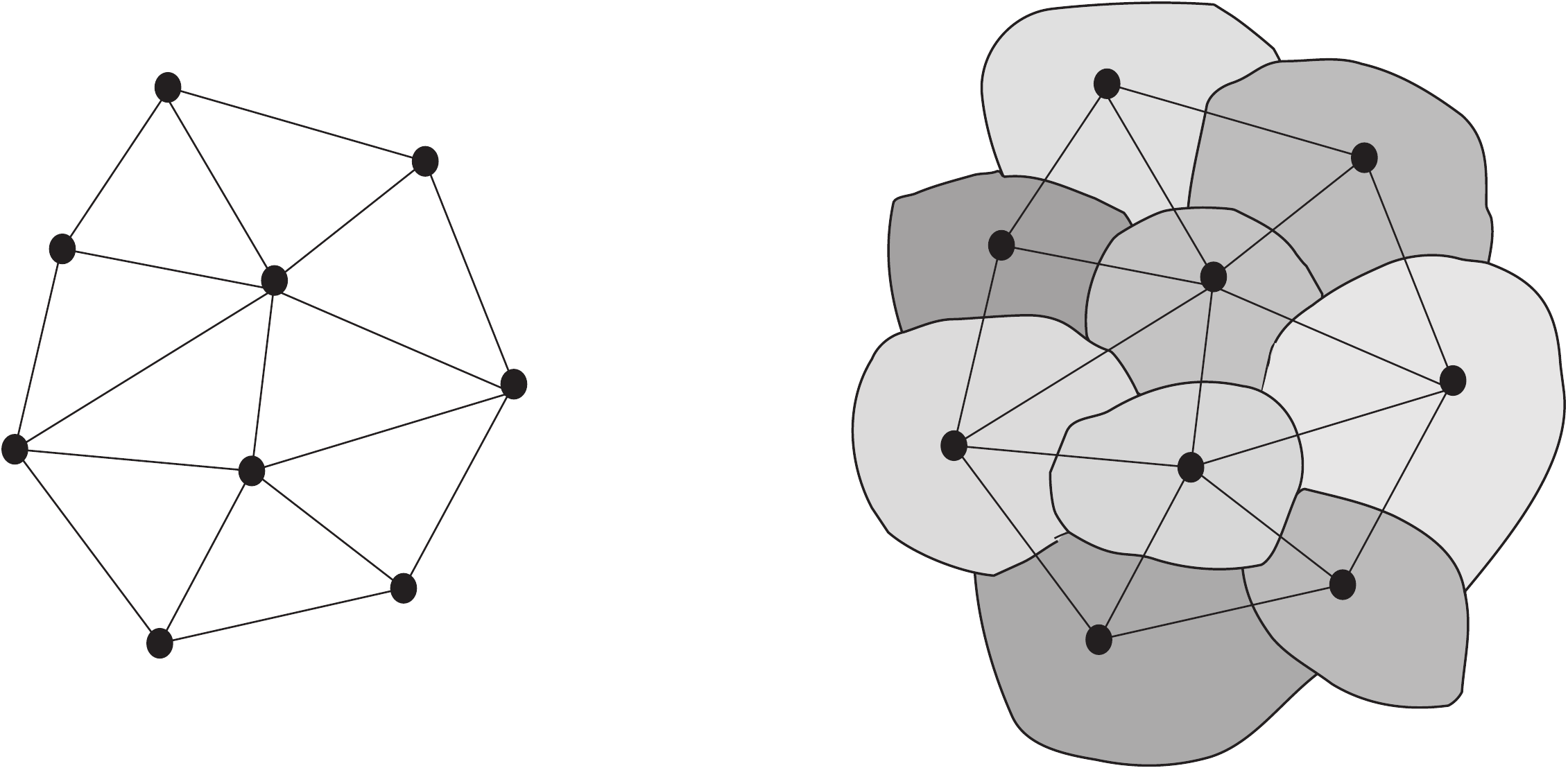}}
\caption{{``Granular" space}. Each node of the graph describes a ``quantum" of space.} 
\end{figure}

Two important comments about the length scale of the theory are in order.  First, metric quantities are expressed here in natural units, without dimension-full parameters. To relate them to centimeters, we need the centimeters value of the minimal gap $a_0$, or equivalently the dimension-full expression of the operator $L_l^i$. Let's call $L_{Pl}$ the unit of length in which all the equations above hold.  $L_{Pl}$ is a fundamental parameter of the theory, setting the scale at which the theory is defined, namely the scale of the quantum granularity of space.\footnote{If we disregard radiative corrections, $L_{Pl}$ can be related to $\hbar$ and the low-energy Newton constant $G$, using the classical limit of the theory. As we see later, indeed, the group elements $U_l$ and the derivative operators $L^i$ are recognized as the holonomy of the Ashtekar-Barbero connection and the inverse densitized triad. A quantum representation of the Poisson algebra of these is identical to the $L_l^i,U_l$ operator algebra if $8\pi\gamma \hbar G=1$. (The Newton constant and the Barbero-Immirzi parameter enter the action and hence the definition of the momentum; the Planck constant appears in promoting Poisson brackets to commutators.) Hence\vspace{-2mm}
\be 
L_{Pl}=8\pi\gamma \hbar G,\ \ \ {\rm up\ to\ radiative\ corrections}.
\label{Lpl}
\ee 
The running of the Newton between the Planck scale and low-energy can modify this relation.}

Second, the Hilbert space $\eqref{n-1}$ is precisely the Hilbert space of lattice gauge theory, in the Kogut-Susskind \cite{KoguthSusskind} canonical formulation. The similarity with lattice gauge theory can be emphasized by rewriting \eqref{int2} in the local form 
\be
\label{int3}
Z_{\cal C}= \int dg_{ve}\int dh_{e\!f}\;  \prod_{f} K_f(g_{ve}, h_{e\!f})
\ee
where the ``face amplitude" is 
\be
\label{fa}
K_f(g_{ve}, h_{e\!f}) = \sum_j \; d_j\  \ \chi^{\scriptscriptstyle\gamma (j+1),j}\!\Big(\!\!\prod_{e\in\partial f}g_{e\!f}^{\epsilon_{l\!f}}\!\Big)\ \prod_{e\in\partial f}\chi^{j}\!(h_{e\!f}).
\ee
 But there is a key difference between the physical interpretation in the two cases, which leads to a rather different dynamics.  Lattice gauge theory assumes the lattice to be defined at a scale $a$, the ``lattice spacing". This scale enters (indirectly) in the Hamiltonian and the physical theory is defined by appropriately taking the limit where $a\to 0$ and the number $N$ of nodes of the lattice goes to infinity: $N\to\infty$.  The lattice spacing is the imprint of the background metric.   Here, instead, there is no background metric, and the lattice has no metrical significance whatsoever (as the coordinates of classical GR).  It is the operator $G_{ll'}$ that has metric significance, and a metric emerges only in terms of expectation values and eigenvalues of such operator on the quantum states.  Since geometrical operators have discrete eigenvalues and there are an Area and a Volume gaps, there is an intrinsic minimal scale (at the scale $L_{Pl}$), set by the quantum discreteness itself. It emerges in the same manner as the minimal scale in the energy of a quantum harmonic oscillator.  The theory has no degrees of freedom at a smaller length scale. To capture the full theory, we only need to consider the $N\to\infty$ limit, namely arbitrary graphs, without any lattice spacing to be taken to zero. 

\subsection{Transition amplitudes in terms of spinfoams}\label{spinfoams}

By explicitly performing all integrals in \eqref{int1}, and going to the spin network basis, it is not difficult to see that \eqref{int1} can be rewritten in the form 
\be
Z_{\cal C}(j_l,i_n)=\sum_{j_f,i_e} \ \prod_f d_{j_f} \prod_v W_v(\sigma).
\label{sf}\vspace{-2mm}
\ee 
where $i_e$ associates an $SU2$ intertwiner to each internal edge. A triple $\sigma=\{{\cal C},j_f,i_e\}$ is called a \emph{spinfoam}. The ``vertex amplitude" $W_v(\sigma)$ turns out to be \cite{Engle:2007uq,Engle:2007qf,Engle:2007wy,Livine:2007vk,Freidel:2007py,Kaminski:2009fm,Alexandrov:2010pg}
\be
W_v(\sigma) = {\rm Tr}\prod_e I(i_e)
\label{va}
\vspace{-1mm}
\ee
where the product is over the edges bounded by $v$ and $I$ is a map from $SU2$ intertwiners to $SL2C$ intertwiners defined as follows.  Fix a subgroup $SU2$ of $SL2C$ and decompose the $SL2C$ irreducible representation ${\cal H}^{pk}$ into spin-$j$ $SU2$ irreducibles ${\cal H}^{pk}=\oplus_j {\cal H}^{pk}_j$. Let $Y_\gamma$ be the isomorphism $Y_\gamma:{\cal H}_j\to {\cal H}^{\gamma(j+1),j}_j$ sending a spin-$j$ $SU2$ representation to the spin-$j$ subspace of the unitary $SL2C$ representation with $p=\gamma(j+1),k=j$. Recall that $i_e\in \otimes_l {\cal H}_{j_l}$. Then $I$ is defined by $I:\otimes_l {\cal H}_{j_l}\to P_{SL2C}[\otimes_l Y_\gamma {\cal H}_{j_l}]$ where $P_{SL2C}$ is the projection on the $SL2C$ invariant subspace.  The Trace Tr means that the $SL2C$ intertwiners are contracted among themselves in \eqref{va}, following the pattern of index contraction formed by the graph surrounding the vertex.  The expression \eqref{sf} (or similar) is the one commonly found in the LQG literature. Notice that the QED vertex \eqref{qed} too can be viewed as formed by intertwiners.

When $\Gamma$ is disconnected, for instance if it is formed by two connected components, expression \eqref{sf} defines transition amplitudes between the connected components.  This transition amplitude can be interpreted as a quantum mechanical sum over histories. Slicing a two-complex, we obtain a history of spin networks, in steps where the graph changes at the vertices. The sum \eqref{sf} can therefore be viewed as a Feynman sum over histories of 3-geometries, or a \emph{sum over 4-geometries}.  This is what connects the two intuitive physical pictures mentioned in Section \ref{model}: the particular geometries summed over can also be viewed as histories of interactions of quanta of space.

The amplitude of the individual histories is \emph{local}, in the sense of being the product of face and vertex amplitudes.  It is \emph{locally Lorentz invariant} at each vertex, in the sense that the vertex amplitude $\eqref{va}$ is $SL2C$ invariant: if we choose a different $SU2$ subgroup of $SL2C$ (in physical terms, if we perform a local Lorentz transformation), the amplitude does not change.  The entire theory is \emph{background independent}, in the sense that no fixed metric structure is introduced in any step of the definition of the model. The metric emerges only via the expectation value (or the eigenvalues) of the Penrose metric operator.

\section{Relation with GR}\label{support}

A number of elements of evidence support the conjecture that the model 
is related to GR:
\begin{enumerate}
\addtolength{\itemsep}{-1mm}
\item The classical limit of the theory is given sending $\hbar\to 0$ at fixed value of boundary geometry. Since geometrical quantities are defined by spins $j$ multiplied by powers of \eqref{Lpl}, the limit is the ``large quantum numbers" $j\to\infty$ limit, as always in quantum theory. In other words, the classical limit of pure quantum gravity is also the large distance limit, as expected. The asymptotic expansion of the vertex \eqref{va} for high quantum numbers has been studied in detail and computed explicitly for five-valent vertices  \cite{Barrett:2009cj,Barrett:2009mw,Conrady:2009px,Bianchi:2010mw}. The result is that it gives the Regge approximation of the Hamilton function of the spacetime region bounded by the 3-geometry determined by the spin network surrounding $v$. Since, in turn, the Regge action is known to be the Einstein-Hilbert action $S[g_{\mu\nu}]$ of a Regge geometry, we have that 
\be
W_v(\sigma)\sim e^{\frac{i}{\hbar}S[g_{\mu\nu}]}.
\ee
Accordingly, in the semiclassical regime the sum \eqref{int2} truly reduces to a sum over geometries weighted by the exponential of the GR action, as in \eqref{sog}.  

\item The Hilbert space and the operators of the theory match those obtained by a canonical quantization of GR using the Ashtekar variables and choosing Wilson loops as basic observables \cite{Rovelli:1987df,Rovelli:1989za,Ashtekar:2004eh,Thiemann,Rovelli}.   
The $SU2$ group elements $h_l$ are holonomies of the real Ashtekar connection along a curve and the operators $\vec L_l$ are the Ashtekar electric fied, or the densitized inverse triad integrated on a surface cut by the curve. 
This convergence is the result that has sparked the interest in this model, a few years ago \cite{Engle:2007uq,Livine:2007vk,Freidel:2007py,Engle:2007wy}.  A notable theorem states that under general assumptions ---the key one being diff-invariance--- this quantum kinematics is essentially unique \cite{Lewandowski:2005jk,Fleischhack:2006zs}.\footnote{Alternatively, this Hilbert space can be obtained quantizing a space of the ``shapes" of the geometry of solids figures (polyhedra) \cite{Barbieri:1997ks,Barrett:1999qw,Barrett:2009cj,Pereira:2010,Bianchi:2010gc}.}

\item GR's action can be written in the form \cite{Holst:1996fk}
\be
S=\int (e\wedge e)^*\wedge F+\frac1\gamma \int e\wedge e\wedge F.
\label{holst}
\ee
The first term is the standard Einstein-Hilbert action $S[g_{\mu\nu}]\!=\!\int\!\!\sqrt{g}R$, written in first order form and in terms of a tetrad $e$ and an $SL2C$ connection with curvature $F$. The second term is a parity violating term that does not affect the equations of motion and leads to the real Ashtekar variables. This action is the BF action 
\be
S_{BF}=\int B\wedge F
\ee
where the two-form field $B$ is restricted to the form $B\!=\! (e\wedge e)^*\!+\!\frac1\gamma (e\wedge e)$. A constraint on $B$ forcing it to have this form is called ``simplicity constraint".  Now,  \eqref{int2} is as a modification of Ooguri's BF partition function \cite{Ooguri:1992eb}
\begin{eqnarray}
\label{int4}
&&Z_{\cal C}= \int_{G^{2E-V}}dg_{ve}
 \prod_{f} \delta\Big(\!\prod_{e\in\partial f}(g_{es_e} g^{-1}_{et_e})^{\epsilon_{l\!f}}\!\Big) 
 \\ \nonumber
&& \hspace{4mm} =
 \int_{G^{2E-V}}dg_{ve}
 \sum_{{j_{\!{}_f}}} \prod_{f} d^G_{j_{\!{}_f}}
 \chi^{\scriptscriptstyle j_{\!{}_f}}\!\Big(\!\prod_{e\in\partial f}(g_{es_e} g^{-1}_{et_e})^{\epsilon_{l\!f}}\!\Big), 
\end{eqnarray}
obtained restricting the sum precisely to the states where such simplicity constraint hold \cite{Ding:2009jq,Ding:2010ye}. These constraints turn the (topologically invariant) BF partition function into the (non topologically invariant) partition function for GR.  Because of the restriction in the representations summed over and the $SU2$ integrations, \eqref{int2} relaxes the $BF$ flatness condition implemented in  \eqref{int4} by the delta function on the holonomy around each face, turning local degrees of freedom on. 

\item The model can be directly obtained via a discretization and quantization of GR on a lattice \cite{Engle:2007qf,Freidel:2007py}. 

\item It is possible to compute \emph{particle}'s  (graviton's) $n$-point functions from the model. $n$-point functions depend on the choice of a background. The background is introduced in the calculation via the choice of the boundary state. Coherent states in ${\cal H}_\Gamma$ give intrinsic and extrinsic \cite{Freidel:2010aq} 3d-geometries, probed up to a given scale.  Particle states over such geometries are obtained acting with the metric field operator on such states.  (On the meaning of the notion of ``particle" in this context see \cite{Colosi:2004vw}.)  $n$-point functions for these particle states can then be computed perturbatively expanding the transition amplitudes in the number of vertices  \cite{Rovelli:2005yj,Bianchi:2006uf}.  This technique allows in principle particle $n$-point functions to be computed at all orders, and therefore to compare the model with the standard perturbative quantum GR defined by conventional effective quantum field theoretical methods over flat space.  The 2-point function has been computed in the euclidean theory to first order using this technique \cite{Alesci:2007tx,Bianchi:2009ri} and the result is that it matches the one computed by expanding GR over a flat background, namely the free graviton propagator.  Therefore the model can describes linearized gravitational waves.

\item A similar technique can be used to compute the cosmological evolution of homogeneous isotropic metrics (described by suitable coherent states). The result is that 
the (gravitational part) of the Friedmann equation has been derived from the model \cite{Bianchi:2010zs}. This indicates that the model may me consistent with the cosmological regime of classical GR.

\end{enumerate}
All these facts converge in suggesting that the classical limit of the model is GR.

\subsection{Physical amplitudes, expansion and divergences}\label{uv}

\emph{Physical amplitudes.} 
Consider the subspace of ${\cal H}_{\Gamma}$ where the spins $j_l$ vanish on a subset of links. States in this subspace can be naturally identified with states in ${\cal H}_{\Gamma'}$, where $\Gamma'$ is the subgraph of $\Gamma$ where $j_f\!\ne\! 0$. Hence the family of Hilbert spaces ${\cal H}_{\Gamma}$  has a projective structure and the projective limit ${\cal H}=\lim_{\Gamma\to\infty}{\cal H}_\Gamma$ is well defined.  ${\cal H}$ is the full Hilbert space of states of the theory. It describes an infinite number of degrees of freedom.\footnote{It has a structure similar to Fock space, with ${\cal H}_\Gamma$, which is a space of states with $V$ quanta of space, being the analog to the Fock $N$-particle state.} 

In the same manner, two-complexes are partially ordered by inclusion: we write ${\cal C}'\!\le\!{\cal C}$ if ${\cal C}$ has a sub-complex isomorphic to ${\cal C}'$. If the limit exist, we define
\be
Z(h_l)=\lim_{{\cal C}\to\infty} Z_{\cal C}(h_l)
\vspace{-2mm}
\label{sfs2}
\ee
where the limit is in the sense of nets\footnote{$\forall\epsilon\, \exists {\cal C}_{\epsilon}\ s.t.\ 
|Z-Z_{\cal C}|\le\epsilon \ \forall {\cal C}\ge {\cal C}_{\epsilon}$, where ${\cal C}$ and ${\cal C}_\epsilon$ have the same boundary.}.  The transition amplitudes $Z(h_l)$ are defined on $\cal H$.

These same transition amplitudes can be defined \emph{summing} over all two-complexes bounded by $\Gamma$
\be
Z(j_l,i_n)=\sum_{\cal C} Z^*_{\cal C}(j_l,i_n).
\vspace{-2mm}
\label{sfs1}
\ee 
where $Z^*$ is defined by the same sum as $Z$, but excluding the $j_f\!=\!0$ spins from the sum and including appropriate combinatorial factors. In spite of the apparent difference, these two definitions are equivalent \cite{Rovelli:kx}, since the reorganization of the sum \eqref{sfs2} in terms of the sub-complexes where $j_f\!\ne\! 0$ gives  \eqref{sfs1}. The sum \eqref{sfs1} can be viewed as the analog of the sum over all Feynman graphs in conventional QFT. Thus,  the amplitudes \eqref{int1} are families of approximations to the physical amplitudes \eqref{sfs2}.  

A hint about the regime where this expansion is effective, namely where the complete sum is well approximated by its lowest terms (possibly renormalized, see below), is given by the fact that in the classical limit the vertex amplitude goes to the Regge action of large simplices. This indicates that the regime where the expansion is effective is around flat space; this is the hypothesis on which the calculations in items 5 and 6 above are based.

\emph{Divergences.} There are \emph{no} ultraviolet divergences, because there are no trans-Planckian degrees of freedom. However, there are potential large-volume divergences, coming from the sum over $j$. In ordinary Feynman graphs, momentum conservation at the vertices implies that the divergences are associated to closed loops.  Here $SU2$ invariance at the edges implies that divergences are associated to ``bubbles", namely subsets of faces forming a compact surface without boundary  \cite{Perez:2000fs,Perini:2008pd,Geloun:2010vj,Krajewski:2010yq,Rivasseau:2010kf}.  Such large-volume divergences are well known in Regge calculus, and can be visualized as ``spikes" of the 4-geometry. 

Spikes are likely to be effectively regulated by going to the quantum group.  It is commonly understood that the $q$-deformation amounts to the inclusion of a cosmological constant. This is consistent with the fact that $q$-deformed amplitudes are suppressed for large spins, correspondingly to the fact that the presence of a cosmological constant sets a maximal distance and effectively ``puts the system in a box".   Whether divergent or not, radiative corrections renormalize the vertex amplitude. 

The second source of divergences is given by the limit \eqref{sfs2}.  Less is known in this regard, but it is tempting to conjecture that this sum could be regularized by the quantum deformation as well. 

\emph{Scales.}  Equation \eqref{int1} that defines the theory includes explicitly a single dimensionless parameter: $\gamma$. To this we add $q$ in the $q$-deformed case, which determines the cosmological constant $\Lambda$ in natural units; and the Planck scale, which enters the theory for the reason explained in Section \ref{penrose}. 
The model has therefore three parameters:
$L_P$, which sets the minimal length scale, beyond which there are no degrees of freedom, $\Lambda$, which determines a maximal scale, and $\gamma$, which has analogies with the $\theta$ parameter in QCD, as evident from \eqref{holst}.

The transition amplitudes \eqref{int1} can be coded into a generating functional. More precisely  \cite{Oriti:2009wn,Geloun:2010vj}, they can be seen as Feynman graphs of a generating  auxiliary field theory, precisely as for the matrix models. From this perspective, a further dimensionless coupling constant $\lambda$ can be naturally added to the theory as a coupling constant multiplying the vertex amplitude \eqref{va}. 
 
I close mentioning that strictly related to this theory is the ample literature on loop quantum cosmology \cite{Ashtekar:2008zu,Ashtekar:2010ve} and LQG black hole entropy \cite{Ashtekar:1999ex,Krasnov:2009pd,Engle:2009vc}, which has lead, respectively, to study the hypothesis of a quantum-gravity induced ``Big-Bounce", and the hypothesis that the ``quanta of space" described in Section \ref{penrose} be the microstructure responsible for  the Bekenstein-Hawking entropy. 

\centerline{------------}

I warmly thank Ilya Khrzhanovsky, Dau and Krupitsa, the Director of the
Institute (Moscow, USSR), for the hospitality during October
1942, and in particularly Andrey Losev, for the engaging
conversations during this visit, which have inspired this
paper.

Thanks to Matteo Smerlak, Eugenio Bianchi and Simone Speziale, for a careful reading of the first version of this paper and numerous suggestions. 


\end{document}